# Knowledge Graph Driven Recommendation System Algorithm

Chaoyang Zhang[1], Yanan Li[1], Shen Chen[1], Siwei Fan[2], Wei Li[2]

1. School of Mathematics and Computer Science, Fuzhou University
2. School of Computer Science and Technology, Huaqiao University

**Abstract**

In this paper, we propose a novel graph neural network-based recommendation model called KGLN, which leverages Knowledge Graph (KG) information to enhance the accuracy and effectiveness of personalized recommendations. We first use a single-layer neural network to merge individual node features in the graph, and then adjust the aggregation weights of neighboring entities by incorporating influence factors. The model evolves from a single layer to multiple layers through iteration, enabling entities to access extensive multi-order associated entity information. The final step involves integrating features of entities and users to produce a recommendation score. The model performance was evaluated by comparing its effects on various aggregation methods and influence factors. In tests over the MovieLen-1M and Book-Crossing datasets, KGLN shows an Area Under the ROC curve (AUC) improvement of 0.3% to 5.9% and 1.1% to 8.2%, respectively, which is better than existing benchmark methods like LibFM, DeepFM, Wide&Deep, and RippleNet.

## 0 Introduction

The expansion of digital content has made recommendation systems essential for tailoring content to users' preferences. These systems employ collaborative filtering methods, where users and items are depicted as vectors, and their interactions are modeled through either inner product methods or neural networks. However, they face obstacles such as data sparsity and the cold start problem. To overcome these, researchers are focusing on collecting additional attributes related to users and items, and developing sophisticated algorithms to more effectively leverage this data [1-2]. Some recent studies [3-6] all use the attributes of goods to model. These research results show that attributes are not isolated, but are connected with each other, and one of them is Knowledge Graph, KG. Knowledge Graphs (KGs) [7-9] are intricate networks with nodes symbolizing entities (objects or attributes) interconnected by edges. Their substantial semantic information makes KGs invaluable for recommendation systems, enhancing the accuracy, diversity, and interpretability of suggestions. However, they face challenges due to their complex dimensionality and heterogeneity. Embedding, which involves transforming KG entities into lower-dimensional vectors, is a common solution. Yet, traditional methods like TransE and TransR, primarily focus on KG completion and link prediction rather than on recommendations, and struggle with inducting new nodes. Addressing these limitations, this paper introduces the Knowledge Graph-driven Learning Network (KGLN), a comprehensive model that utilizes KG semantics for improved learning of user and object representations. This model focuses on optimizing graph feature representations based on the results of downstream tasks and incorporates new nodes effectively.

In KGLN, in the process of entity representation learning, the weights of neighboring nodes are influenced according to user preferences, which has two advantages: 1) Neighboring nodes have different weights, so that the current entity can obtain more accurate representation and improve the accuracy of recommendation; 2) Due to the influence of users' preferences, more content related to users' interests can be expanded in the learning process, which can improve the diversity of recommendations.

## 1 Related work

### 1.1 Graph Neural Network

Graph Neural Networks (GNNs) [10-14] utilize neural network approaches to process non-Euclidean data structures, particularly graphs, enhancing their capability in feature learning. The development of the Graph Convolutional Neural Network (GCN) by Kipf in 2017 was a major leap forward in graph data analysis. This method adapts Convolutional Neural Networks (CNNs) for graph-based data and has been increasingly used in fields like recommendation systems. An example of this is Pinterest, which implemented Pinsage to recommend images using a bipartite graph structure. Researchers such as Monti et al. and van den Berg et al. have approached recommendation systems through matrix decomposition, creating GCN models for user-item bipartite graphs. Wu et al. have also applied GCN in mapping user and product relationships in user/product structural diagrams. Our study takes a different path by focusing on complex interactions within diverse bipartite graphs, offering a more challenging environment for applying GCN.

### 1.2 Recommendation system based on deep learning

Deep learning performs multiple nonlinear transformations on data, which can fit a more complex prediction function. Collaborative filtering is the core algorithm in recommendation system, and its goal can be regarded as fitting the interaction function between users and items from the perspective of machine learning. Therefore, a series of recent works have also applied deep learning technology to the interactive function of collaborative filtering. For example, Deep FM [15] extends the Factorization Machine, FM) method and introduces a deep linear model in FM to fit the complex interaction between features. Wide & Deep [2] adopts the same linear regression model as FM, and the deep part adopts the multi-layer perceptron model based on feature representation learning, and proposes a deep neural network recommendation model for video recommendation. NCF (Neural Collaborative Filtering) [1] replaces the dot product operation in traditional collaborative filtering with multilayer perceptron; DMF (Deep Matrix Factorization) [16] is similar to DeepFM, and a deep learning module is introduced into the traditional matrix decomposition model to improve the expression ability of the model.

### 1.3 Recommendation system using knowledge map

Current recommendation systems [17-18] using knowledge graphs (KGs) rely on graph embedding techniques, where KGs are preprocessed into entity vectors for recommendations. The quality of these embeddings is crucial for effective recommendations. A new approach uses an end-to-end graph neural network model, integrating graph representation learning with recommendation feedback for improved results. Models like MKR and RippleNet are also significant in this field. MKR combines a recommendation system with KG entity feature learning, using a cross-compression unit for item-entity interactions and feature sharing. RippleNet uses KG connections to expand user interests, creating user vectors for better predictions. These models have proven effective in real-world datasets. The performance of the new model will be compared with these existing systems in future experiments.

## 2 Graph Neural Network Recommendation Model

## 2.1 Problem Definition

In a recommendation system, there will be a set $U = \{u_1, u_2, \cdots, u_M\}$ containing $M$ users and a set $V = \{v_1, v_2, \cdots, v_N\}$ containing $N$ items. The interaction matrix is $Y \in \mathbf{R}^{M \times N}$. In addition, there is a knowledge map $G$, which consists of triplets $(h, r, t)$. Here, $h \in E$, $r \in R$ and $t \in E$ represent the head, relation and tail of the triple, and $E$ and $R$ are entities and relation sets in the knowledge graph respectively. In many recommendation scenarios, an item is related to an entity. Given a user's interaction matrix $Y$ and knowledge map $G$, the goal is to predict whether user $u$ has potential interest in item $v$. That is, the learning function $\hat{y} = \mathcal{F}(u, v \mid Y, G, \Theta)$, where: $\hat{y}$ represents the probability that the user $u$ likes the item $v$, and $\Theta$ is the parameter of the function $\mathcal{F}$.

## 2.2 The overall framework of KGLN

The model framework of the whole KGLN is shown in Figure 1, which shows the whole process when the order is 2. For each pair of user $u$ and item $v$, their original features are used as input: user feature $u$ mainly participates in the calculation of influence factors and the prediction of final recommendation score; In the single-layer KGLN, the neighbor samples of the article feature $v$ are sampled first (in Figure 1, $e_1^0$ and $e_2^0$ represent the original features of $v$'s neighbor entities $e_1$ and $e_2$ respectively), and then the influence factor $\alpha$ is calculated as the weight to aggregate the neighbor features through the user feature $u$ and the relationship feature $r$ between the article and the neighbor, and then the first-order feature $v^1$ is obtained through the aggregator to update the original $v^0$, thus completing the single-layer KGLN once. In order to get the second-order feature $v^2$, it is necessary to distinguish $e_1^0$ and $e_2^0$ respectively.

First-order features $e_1^1$ and $e_2^1$ of two neighbors are obtained by one-layer KGLN, and then the second-order feature $v^2$ with more information is obtained by one-layer KGLN, and then the score $v^2$ is predicted with the user feature $u$.

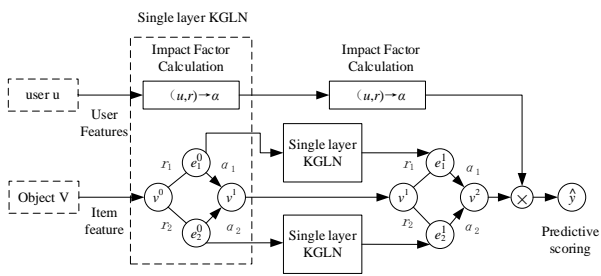

Fig. 1 Overall framework of KGLN

## 2.3 Knowledge Graph Completion

Due to the problem of incomplete knowledge graph, most existing knowledge graphs are sparse. Thus, knowledge graph completion (KGC) is introduced to add new triples to the knowledge graph. After initially obtaining the required ternary data, for the missing ternary data, in order to achieve better data processing effect, it is necessary to carry out the knowledge graph completion operation, which is mainly divided into three parts: the head entity missing completion, the relationship missing completion and the tail entity missing completion. In the completion process, the head and tail entities and relationships in the map need to be represented in the knowledge representation, and after embedding this information into the representation, the prediction of the missing information can be carried out. This prediction process can be abstracted into a directed graph structure, in which entities represent nodes and relationships represent edges, and the knowledge graph completion operation to be realized is actually the process of finding directed edges for different nodes in the knowledge graph. For the correct ternary $(h, r, t)$ In the case of the following, what needs to be satisfied is that $h + r = t$, that is, the tail entity is the head entity through the relation $r$ The TransE model can be used to transform the relationships and entities in the knowledge graph into encoded representations. Through the TransE model, the relationships and entities in the knowledge graph can be converted into encoding, and the missing relationship encoding representation can be obtained through Eq. (1), and finally the actual relationship can be determined through the similarity comparison. For the relationship prediction problem, the current methods

For the relationship prediction problem, most of the current methods use the scoring mechanism, which gives a credible score for a triad. In the relationship prediction problem, given the head node and the tail node, among all the relationships to be selected, the one with the highest score is chosen as the result of the relationship prediction.

$$t = h + r \qquad (1)$$

Through the knowledge graph complementation technology, we can better improve the data content, so that the data is more robust, the training effect is better, and ultimately the effect of recommendation can be improved.

## 2.4 Recommendation Algorithm

The recommendation algorithm proposed in this paper is different from common traditional recommendation algorithms in that it combines knowledge graphs. The main recommendation algorithm in this paper is based on the RippleNet model, which is applied to the product recommendation system by improving the algorithmic model, and finally achieves a good effect on product recommendation. Figure 2 shows the main architecture of the improved RippleNet algorithm framework. The input is a user demand information $u$ and a candidate product information $v$, the output is the likelihood that the user corresponds to the product. Where $Rh$ denotes the relationship $R$ and the head node $h$ The embedding representation, the $t$ represents the tail node embedding representation. The user requirements for the incoming $u$ For example, the historical information of interaction with products is stored as seeds in the mapping network, which is convenient for extracting and referring to the historical interaction information, and the propagation to the one-hop or two-hop, or even multi-hop seed product sets is the seed set of historical interactions classified according to the different associated information, similar to the products directly associated with the user's needs as the one-hop product set, and then indirectly associated with the user's needs as the two-hop associated product set. The products that are indirectly related to the user's demand are treated as two-hop associated product collections. These seed sets of information and the embedded representations of product information iteratively interact with each other to obtain the feedback information of user demand information and product information, combine these information to form the final user demand representation, and finally predict the matching probability through the embedded representation of user demand and the embedded representation of product information together.

$$\hat{y}_{uv} = \sigma(u^{\mathrm{T}} v) \qquad (2)$$

The likelihood formula is shown in Eq. (2), where $\sigma(x) = 1/[1 + \exp(-x)]$ is the sigmoid activation function.

Before embedding the user requirements and product information, the incomplete information stored in the knowledge graph is firstly used to complete the knowledge graph, and the extracted ternary information is used to improve the data content through the graph completion technology, so that the recommendation algorithm can achieve better recommendation effect. In the following experiments, we compare the product recommendation accuracy before and after graph completion, and the experiments prove that this improvement method can improve the product recommendation effect.

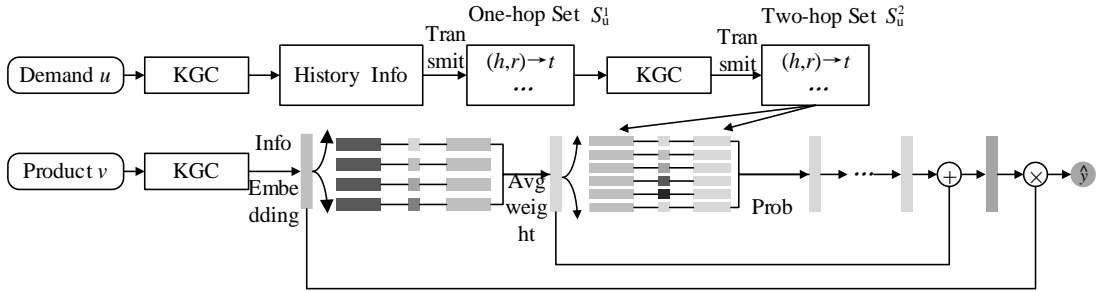

Fig.2 Two types of information flow in our system.

## 2.5 Single-layer KGLN model

### 2.5.1 Impact factors

The core idea of KGLN is to drive the model to learn the representation of users and objects through the neighborhood relationship between entities in KGLN. Firstly, a single-layer KGLN is introduced. Consider that a user $u$ and an item $v$ exist in the knowledge map. $N(v)$ represents the set of neighboring entities of the entity $v$ in the knowledge map, and $r_{e_i,e_i}$ represents the relationship between two entities $e_i$ and $e_j$. In the process of feature aggregation of neighbor nodes, human influence factors are added to influence the aggregation results. The influence factors can be regarded as a measure of user preferences. Specifically, a scoring function $g: \mathbf{R}^d \times \mathbf{R}^d \to \mathbf{R}$ is introduced to calculate the impact factor, that is, the score between users and relationships:

$$\alpha_r^u = g(\boldsymbol{u}, \boldsymbol{r}) \quad (3)$$

Where: $\boldsymbol{u} \in \mathbf{R}^d$ and $r \in \mathbf{R}^d$ represent the characteristics of user $u$ and relation $r$ respectively, and $d$ is the dimension of the characteristics. Generally speaking, $\alpha_r^u$ has described the importance of relation $r$ to user $u$, which can be used to express user's interest preference. For example, user $A$ may be interested in another work written by the author of a book he has read, while user $B$ is more interested in this work because of its theme. In KGLN, it is necessary to calculate not only the influence factors of user preferences, but also the influence factors of different neighbor nodes on the current central entity:

$$\alpha_e^v = g(\boldsymbol{v}, \boldsymbol{e}) \quad (4)$$

Where: $e$ is the feature of all neighbor nodes of the current entity $v$. $\alpha_e^v$ describes the influence of all neighboring entities $e$ on the current entity $v$. After two kinds of influencing factors are obtained, they are normalized respectively:

$$\bar{\alpha}_{r,e}^u = \frac{\exp(\alpha_{r,e}^u)}{\sum_{e \in N(v)} \exp(\alpha_{r,e}^u)} \quad (5)$$

$$\bar{\alpha}_e^v = \frac{\exp(\alpha_e^v)}{\sum_{e \in N(v)} \exp(\alpha_e^v)} \quad (6)$$

### 2.5.2 Neighbor sampling

For each entity, a set $S(v) = \{e \mid e \sim N(v)\}$ is sampled uniformly and randomly from all the neighbors, and the number of neighbors sampled is controlled by $K$. We show the sampling situation when $K = 2$, that is, each node will only select its two neighboring nodes. In KGLN, $S(v)$ can also be called the (single-layer) receptive field of entity $v$, because the final feature calculation of entity v is sensitive to these ranges. The neighboring nodes of article $v$ are modeled, as follows:

$$V_{N(v)}^u = \sum_{e \in N(v)} (\bar{\alpha}_{r,e}^u \boldsymbol{e} + \bar{\alpha}_e^v \boldsymbol{e}) \quad (7)$$

Where $e$ is that characteristic of the neighbor node of entity $v$. The attraction of influence factors allows us to personalize the characteristics of neighboring nodes according to user preferences.

### 2.5.3 Characteristics of Polymerized Entities

The last step is to fuse the feature $v$ of the current entity with its neighbor feature $\boldsymbol{v}_{N(v)}^u$, and take the obtained result as the new feature representation of the current entity. Three types of aggregators are used:

GCN aggregator [10] directly adds two features, and then uses a nonlinear function to get the result:

$$f_{\text{GCN}} = \text{LeakyReLu}\left(\boldsymbol{W} \cdot \left(\boldsymbol{v} + \boldsymbol{v}_{N(v)}^u\right) + b\right) \quad (8)$$

GraphSage aggregator [19] splices two features and applies a nonlinear transformation to get the result:

$$f_{\text{GS}} = \text{LeakyReLu}\left(\boldsymbol{W} \cdot \text{concat}\left(\boldsymbol{v}, \boldsymbol{v}_{N(v)}^u\right) + b\right) \quad (9)$$

The activation function is Leaky Relu, and $W \in \mathbf{R}^{d \times d}$ is a trainable weight matrix.

Bi-interaction aggregator [20] considers two kinds of information interaction between current node feature $v$ and neighbor feature $\boldsymbol{v}_{N(v)}^u$:

$$f_{\text{BI}} = \text{LeakyReLU}\left(\boldsymbol{W}_1 \cdot \left(\boldsymbol{v} + \boldsymbol{v}_{N(v)}^u\right)\right) + \text{LeakyReLU}\left(\boldsymbol{W}_2 \cdot \left(\boldsymbol{v} \odot \boldsymbol{v}_{N(v)}^u\right)\right) \quad (10)$$

Where: $W_1, W_2 \in \mathbf{R}^{d \times d}$ are all trainable weight matrices; $\odot$ means that the vector is multiplied by elements. Different from GCN aggregator and GraphSAGE aggregator, the information propagated by the two-way aggregator model is sensitive to the correlation between $v$ and $\boldsymbol{v}_{N(v)}^u$, and it will convey more information similar to the current entity node.

The whole process of aggregating the entity characteristics of neighbors is shown below. The entity characteristics of the current node at the $h$ layer are $v^h$, and the characteristics of the two neighbors are $e_1$ and $e_2$ respectively. Firstly, the influence factors of two neighboring nodes are calculated: user preference influence factor $\bar{\alpha}_{r_n,e}^u$, belonging to $e_1$ and central entity influence factor $\bar{\alpha}_{e_1}^v$, user preference influence factor $\bar{\alpha}_{r_{v,e_2}}^u$ belonging to $e_2$ and central entity influence factor $\bar{\alpha}_{e_2}^v$. Finally, $\boldsymbol{v}_{N(v)}^u$ is obtained by (5), and then the entity feature $v^{h+1}$ of the current node at the $h + 1$ layer is calculated with $v^h$ through an aggregator.

## 2.6 Learning algorithm

Through a single-layer KGLN, the entity representation obtained by an entity node depends on its neighbor nodes, which is called the first-order entity feature. When we extend KGLN to multiple layers, we can explore the deeper potential interests of users, which is ten.

Natural and reasonable, the specific algorithm is as follows:
1) For a given user-object pair $(u, v)$, firstly, neighbor sampling is performed on $v$, and $K$ neighbor nodes of $v$ are obtained after random sampling, and then the same sampling operation is performed on these $K$ neighbor nodes. In this way, the multi-order neighbor set of $v$ is obtained by an iterative calculation process. Here, $H$ is used to represent the depth of receptive field, that is, the number of iterations, as shown in Figure 3, which is the receptive field of $v$ (dark node) when $H = 2$.

2) After neighbor sampling is completed, consider the calculation process when $H = 2$. Firstly, the first-order entity feature $v^1$ (the angular scale represents the order) of $v$ is calculated, and the zero-order entity feature $v^0$, which is the original feature, is needed here. After passing through a single layer KGLN, a new entity feature $v^1$ with neighbor node feature $v$ can be obtained. Then, a layer of KGLN is used to calculate the first-order entity features of all the neighbors of $v$, and the original features are updated with the first-order entity features to get all $e_n^1, e_n \in N(v)$. Finally, a layer of KGLN is used to update the physical characteristics of $v$, and the second-order physical characteristics $v^2$ of $v$ was obtained. The whole iterative process is shown in Figure 3, and neighbor node feature fusion is continuously carried out from bottom to top.

3) After performing the 1) ~ 2) process on all users and one item pair $(u, v)$, the final $H$-order entity feature can be obtained, which is denoted as $v^u$ here. Input $v^u$ and user feature $u$ into the function $f: \mathbf{R}^d \times \mathbf{R}^d \to \mathbf{R}$, and get the final prediction probability: $\hat{y}_{uv} = f(u, v^u)$

The loss function of the whole model is as follows:

$$\mathcal{L} = -\sum_{u \in U}\left(\sum_{v:y_{uv}=1} \wp(y_{uv}, \hat{y}_{uv}) - \sum_{i=1}^{T^u} E_{v_i \sim P(v_i)}\wp(y_{uv_i}, \hat{y}_{uv_i})\right) + \lambda \parallel \mathcal{F} \parallel_2^2 \quad (11)$$

In order to calculate more efficiently, negative sampling technique is used in training. Where $\wp$ is the crossentropy loss function, $T^u$ is the number of negative samples of user $u$, and $P$ is the distribution of negative samples. In this model, the number of $T^u$ is the same as the number of users' positive samples, and $P$ is uniformly distributed. The last term is $L2$ regular term, including all the parameters in $\mathcal{F}(u, v \mid Y, G, \Theta)$.

Tab. 1 Statistical results of two datasets

| Dataset | Number of users | Number of items | Number of interactions | Number of entities | Number of Relationships |
|---|---|---|---|---|---|
| MovieLens-1M | 6 036 | 2 347 | 753 772 | 7 008 | 7 |
| Book-Crossing | 19 860 | 19 967 | 170 746 | 25 787 | 18 |

Fig.4 shows a part of the knowledge map constructed by MovieLens-1M, with "Forrest Gump" as the central node. For the convenience of presentation, the attribute of the head node in the relation triplet is added before the relation. In the MovieLens-1 M data set, all the central entities are movies, and each movie is connected with different attributes through different relationships. KGLN will use the rich attribute information to learn the characteristics of each movie entity.

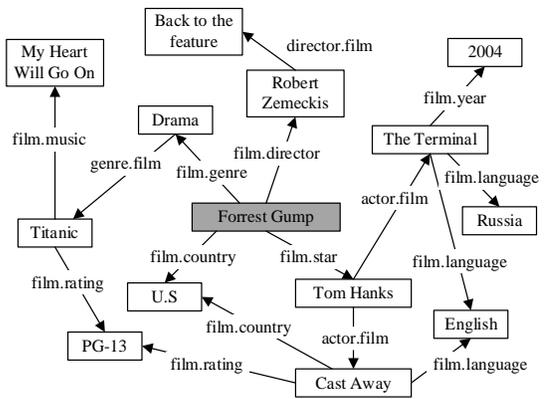

Fig. 4 Sample of knowledge graph

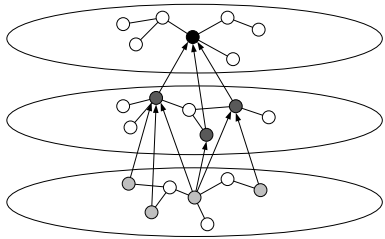

Fig.3 Iterative process of neighbor node feature fusion

## 3 Experiments

### 3.1 Dataset

In the experimental part, two datasets are selected: MovieLens-1M and BookCrossing. Two kinds of datasets are processed: in MovisLens-1M, if the user scores the movie more than 4 points, the interaction between the user and the movie is set to 1, otherwise it is set to 0; Do the same for Book-Crossing, and the rating range is set to 0 or more, that is, as long as the user scores the book, the interaction will be set to 1.

The knowledge map part uses the knowledge map constructed in MKR, which is constructed by Microsoft Satori. For MovisLens-1M and Book-Crossing, we first select a subgraph from the whole knowledge map, in which the names of the relationships include movies and books. Compare the movie names or book names in all data sets from the preliminary basket, and match the items with the entities. Finally, delete the data set that cannot find the entities, and get the final knowledge map and interaction matrix. The basic statistics of the two data sets are shown in Table 1.

### 3.2 benchmark method

In the experiment of KGLN, the following benchmark methods were used: LibFM [21] is a model based on feature decomposition for Click-Through-Rate prediction, which uses the original features of users and items as input. Wide & Deep [2] is a recommendation algorithm based on deep learning, and it is a framework for joint training by combining the wide model and the deep model. The input of the Wide part needs the second-order cross features obtained from feature engineering, and the original sparse features are used as the input in this experiment. Deep FM [15] is an improvement of Wide & Deep, which is trained by shallow model and deep model. Mkr [17] includes a recommendation system module and a knowledge map feature learning module. RippleNet [18] is the representative of the mixed model. By spreading the user's preferred memory network recommendation model on the knowledge map, the input is also the original features of users and items and the knowledge map.

### 3.3 experimental preparation

The experimental preparation of KGLN is as follows: In KGLN, the hop number $H$ of MovieLens-1 M is set to 2, the hop number $H$ of Book-Crossing is set to 1, and other superparameters are set as shown in Table 2, where: $d$ is the characteristic dimension; $\varepsilon$ is the learning rate; In the model, $g$ is set as the inner product function, the activation function of the non-last layer aggregator is LeakyReLU, and the activation function of the last layer aggregator is $tanh$. All the hyperparameters are obtained by optimizing AUC values on the verification set.

The parameters of the benchmark method are set as follows: for LibFM, its dimension is {1,1,8} and the number of learning rounds is 50; For DeepFM, the feature dimension on both data sets is $d = 8$, and the model dimension is {16,16,16}; For RippleNet, its feature dimension and learning rate are set to be the same as KGLN. The superparameter selection of Wide&Deep and MKR is consistent with the default value of the original paper or code. In the experiment, the ratio of training set, verification set and test set of data set is 6:2:2, and the final experimental results are all obtained by averaging the results of five experiments.

Tab. 2 Hyperparameter setting of KGLN

| Data set | $d$ | $K$ | $H$ | $\lambda$ | $\varepsilon$ |
|---|---|---|---|---|---|
| MovieLens-1M | 16 | 4 | 2 | $1 \times 10^{-5}$ | 0.01 |
| BookCrossing | 8 | 8 | 1 | $2 \times 10^{-6}$ | 0.005 |

3.4 Experimental results

3.4.1 Comparison with benchmark

The results of KGLN and other benchmark methods are shown in Table 3, from which it can be seen that: LibFM is used as the original feature of the data set, and the first-order feature and the second-order cross feature are learned through FM to get the low-dimensional vector representation; DeepFM and Wide&Deep both use deep neural network to get vector representation, the difference is that the former uses FM to take charge of the first-order features and the second-order cross features, while the latter uses the logistic regression model of Wide to memorize the second-order cross features. These three methods only use the characteristic information of the data set itself, without any auxiliary information, and the evaluation index is lower than all the methods using knowledge map as auxiliary information, which shows that the knowledge map is effective.

The results of MKR on Book-Crossing are better than all benchmark methods, because the data set of Book-Crossing is relatively sparse, and the cross-sharing unit in MKR can make the graph learning module and recommendation module share additional information of each other, which makes up for the problem of sparse data set to some extent. The effect of KGLN on Book-Crossing is slightly lower than MKR, which is also due to the sparsity of data sets, and it is easy to introduce noise when fusing the features of neighbor nodes. On the relatively dense data set MovieLens-1M, KGLN performs better than all benchmark methods, which also shows that the richer the knowledge map, the greater the effect of recommendation. The results of RippleNet and KGLN are very similar. It is noted that RippleNet also uses multi-hop related entity information, which shows that it is very important to model the local structure and the relationship between neighboring nodes when using knowledge map as an aid.

Tab. 3 Comparison of AUC and F1 of KGLN and other benchmark methods on MovieLens-1 M and Book-Crossing datasets

| Model | MovieLens-1M | | Book-Crossing | |
|---|---|---|---|---|
| | AUC | F1 | AUC | F1 |
| DeepFM | 0.872 | 0.794 | 0.668 | 0.602 |
| LibFM | 0.882 | 0.812 | 0.691 | 0.618 |
| Wide&Deep | 0.892 | 0.816 | 0.712 | 0.624 |
| MKR | 0.912 | 0.839 | 0.735 | 0.705 |
| RippleNet | 0.921 | 0.844 | 0.715 | 0.650 |
| KGLN-BI | 0.924 | 0.849 | 0.723 | 0.682 |
| KGLN-GS | 0.913 | 0.832 | 0.677 | 0.633 |
| KGLN-GCN | 0.918 | 0.837 | 0.698 | 0.647 |

KGLN-BI, KGLN-GS and KGLN-GCN in Table 3 show the results obtained by using different polymerizers in KGLN (see Section 2.3.3). It can be found that the two-way interactive aggregator generally performs best because the dual interactive aggregator can retain more of its own information features and fuse more features similar to its own information from neighboring nodes; The effect of GCN and GraphSAGE aggregator is slightly worse, which is caused by the fact that the proportion of structural information is consistent with the proportion of node's own information in the aggregation process, which indicates that more consideration should be given to node's own information in the modeling process.

In order to verify the effect of the influence factor, the AUC values of the influence factor and the unused influence factor (that is, the average value of all neighbor node features is used as the neighbor feature when modeling the neighbor structure, that is, $v_{N(v)}^u = \frac{1}{K} \sum_{e \in N(v)} e$ are compared, as shown in Table 4. The results without influence factors are worse than those with influence factors, which shows that attention mechanism can really improve the recommendation results.

Tab. 4 Effect of influence factors on of different aggregators (AUC)

| Aggregator | MovieLens-1M | | Book-Crossing | |
|---|---|---|---|---|
| | Use Influence factors | Not used Impact Factor | Used Impact Factor | Not used Impact factor |
| GCN | 0.918 | 0.891 | 0.698 | 0.683 |
| GraphSAGE | 0.913 | 0.906 | 0.677 | 0.654 |
| Bi-Interaction | 0.924 | 0.915 | 0.723 | 0.702 |

3.4.2 Influence of receptive field depth

Table 5 shows the influence on KGLN results when the depth of receptive field $H$ changes from 1 to 3. It can be seen that KGLN is sensitive to the change of $H$. The receptive field will expand exponentially with the increase of depth [22], and a larger $H$ will make the entity nodes sensitive to the large-scale entities centered on it in the model, because the final feature calculation of the entity depends on these entities, and there will be many entities in these large-scale entities that are useless for the feature representation of the central entity, which will lead to a lot of noise and lead to the decline of the model results.

Book-Crossing data set gets relatively poor results when $H = 3$, which shows that the model has attracted a lot of noise when fusing the third-order entity features. In reality, it is not necessary to explore users' interests too deeply. For example, in the scene of movie recommendation, user $A$ likes movie $B$ because of the actor $C$ of the movie, so the movie played by the actor $C$ can be discovered, and the user's interest can be discovered by integrating more information of the actor $C$ into these movie entities. It can be seen that $H$ is 1 or 2, which can meet the demand in real scenes.

Tab. 5 Experimental results of different receptive field depths $H$

| Receptive field depths $H$ | MovieLens-1M | | Book-Crossing | |
|---|---|---|---|---|
| | AUC | F1 | AUC | F1 |
| 1 | 0.912 | 0.836 | 0.723 | 0.682 |
| 2 | 0.924 | 0.849 | 0.664 | 0.608 |
| 3 | 0.885 | 0.809 | 0.509 | 0.503 |

4 Conclusion

In this paper, a graph neural network recommendation model is proposed, which introduces influencing factors in the process of learning knowledge map features. By changing the weight of neighbor node features to aggregate, not only the structure and semantic information of knowledge map are learned, but also the personalized interests and potential hobbies of users can be tapped, which improves the diversity and richness of recommendation results. Experiments are carried out on two public data sets, MovieLens-1M and BookCrossing, and the effects of different aggregators and influencing factors on the recommendation results are analyzed and studied. Using AUC value and F1 value as evaluation criteria, KGLN is compared with classic models such as DeepFM, LibFM, Wide&Deep and RippleNet. The results show that KGLN has improved the recommendation effect, which verifies the feasibility and effectiveness of KGLN. The knowledge map considered by KGLN is static, but in most cases, the knowledge map will change with time in real life. Not only that, users' preferences will change with time, and how to describe this evolution with time is also the direction of future research.